# Bridging Fidelities to Predict Nanoindentation Tip Radii Using Interpretable Deep Learning Models

CLAUS O.W. TROST,[1] STANISLAV ZAK,[1] SEBASTIAN SCHAFFER,[2,3] CHRISTIAN SARINGER,[4] LUKAS EXL,[2,3] and MEGAN J. CORDILL [1,5,6]

1.—Erich Schmid Institute of Materials Science, Austrian Academy of Sciences, Jahnstrasse 12, 8700 Leoben, Austria. 2.—Wolfgang Pauli Institute c/o Faculty of Mathematics, University of Vienna, Oskar-Morgenstern-Platz 1, 1090 Vienna, Austria. 3.—University of Vienna Research Platform MMM Mathematics - Magnetism - Materials, University of Vienna, Oskar-Morgenstern-Platz 1, 1090 Vienna, Austria. 4.—Christian Doppler Laboratory for Advanced Coated Cutting Tools at the Department of Materials Science, Montanuniversität Leoben, Franz-Josef-Straße 18, 8700 Leoben, Austria. 5.—Department of Materials Science, Montanuniversität Leoben, Franz-Josef-Straße 18, 8700 Leoben, Austria. 6.—e-mail: megan.cordill@oeaw.ac.at

As the need for miniaturized structural and functional materials has increased, the need for precise materials characterizaton has also expanded. Nanoindentation is a popular method that can be used to measure material mechanical behavior which enables high-throughput experiments and, in some cases, can also provide images of the indented area through scanning. Both indenting and scanning can cause tip wear that can influence the measurements. Therefore, precise characterization of tip radii is needed to improve data evaluation. A data fusion method is introduced which uses finite element simulations and experimental data to estimate the tip radius in situ in a meaningful way using an interpretable multi-fidelity deep learning approach. By interpreting the machine learning models, it is shown that the approaches are able to accurately capture physical indentation phenomena.

## INTRODUCTION

Nanoindentation has become a standard way to characterize the mechanical properties of a wide range of materials. By being nearly non-destructive and fast to apply, it has advantages over other standard characterization methods, such as tensile testing. Sharp indentation tips are often produced as 3-sided pyramids of Berkovich or cube corner geometry, and both tip shapes are widely used in materials research. Load–displacement ($P$-$h$) indentation curves are used to calculate the elastic modulus and hardness of a material following a procedure described by Oliver and Pharr.[1] To further develop the indentation experiments and to enable the predictions of elastic–plastic material properties from indentation data, various methods have been proposed. Dimensional analysis (Π-Theorem),[2–6] optimization algorithms,[7,8] databases,[9] and different machine learning approaches[10–12] have been used to interpret the data and give insights beyond Oliver and Pharr's analysis. These approaches usually utilize finite element (FE) calculations as the basis of the models. Different material models (elastic–perfectly plastic,[4] power law hardening,[2,6,7,12,13] combined isotropic/kinematic hardening,[14] and Ludwik-type isotropic hardening)[8] have been studied to describe the behavior of indented materials. For the sake of simplicity and to save calculation time, 2D simulations of a cone with 70.3° angle, resembling a Berkovich tip in 2D have often been used. More calculation time consuming 3D simulations have also been used.[10,13–15] Various authors have concluded that 2D and 3D simulations do not yield equivalent results for both sharp[13,16] and blunt tips[14,15] for different material models. Despite this, 2D simulations are still used, since they have a clear advantage in reduced calculation time. Lu et al.[10] proposed a multi-fidelity approach, based on the results of Ref. 17, to combine 2D and 3D simulations with experimental data in one model, thereby, exploiting the advantages of each fidelity, while minimizing the need of actual experimental data to predict elastic–plastic material properties.[10]

When nanoindentation experiments are simulated, a perfectly sharp indenter shape is often assumed.[2,5,7] However, since shape imperfections affect actual measurements, leading to a high level of uncertainties, they have to be accounted for Ref. 15. The radius of the indenter tip usually lies between 50 nm and 100 nm for new indenters, but can be as blunt as 1 $\mu$m due to extended use. Solving the inverse problem $\Pi$-Theorem for a blunt tip geometry becomes more complex, as described by Cheng et al.[4] They pointed out that introducing the tip radius breaks the well-known Kick's law[18] ($h^2$ dependence, with $h$ being displacement) of the loading curve.[4] Therefore, linear and constant terms are added to capture the behavior introduced by the tip radius.[4] Li et al.[8] showed that, by analyzing normalized indentation curves of 2D models, the tip radius can be estimated using parameters extracted from these curves. Since the influences of the tip radius are important, especially in real experiments, monitoring the tip radius directly from the obtained data would not only enable new insight into tip wear effects but also improve the overall data analysis. For this reason, multiple new studies have been carried out with the goal of characterizing tip wear.[15,19–21]

Here, a deep learning approach will be used to estimate the tip radius of experimental nanoindentation data. It will be shown that rather complex machine learning models are still able to capture the physics, and that it is crucial to have the ability to explain such models. Interpretation of machine learning models is already widely used in areas involving serious risks, such as in the field of medicine.[22] In materials science, it is also important that methods are interpretable to ensure their scientific understanding and wide acceptance.[23] The possibility of explaining machine learning results will further accelerate the shift towards data-driven materials science. More can be understood about the operations inside the machine learning "black boxes" by using a cooperative game theory-based approach. SHAP (Shapley Additive exPlanation) is based on Shapley values,[24] and its strength is that it enables a model's agnostic interpretation of individual input values (features) in a detailed, consistent, and local manner.[25] Recently, SHAP has been used in materials discovery and property prediction,[26–31] but not in the field of materials mechanics, as presented here.

The main focus of this study will be on the determination of the tip radii of both FE simulations and experiments. For this reason, 2D and 3D Abaqus FE simulations were created with tip radii ranging from 50 nm to 1 $\mu$m. Features from the FE and experimental indentation curve and its normalized counterpart were used to interpret the data similar to Ref. 10. Tip radii were directly evaluated using a novel self-imaging method described in Ref. 19. Direct tip radius estimations are expected to improve high-throughput indentation experiments, by improving data evaluation quality.

## METHODS

### Simulations

All simulations were performed using the professional FE code Abaqus CAE 2019 with Python scripting to loop both materials parameters and tip geometry. Two models were produced: a 2D axisymmetric model of a cone-shaped indenter, and a 3D model of the Berkovich indenter, taking advantage of the 1/6 symmetry of the tip geometry (Fig. 1a) to minimize the model complexity. The tip radius was introduced by cutting out a sharp point of the indenter tip by revolving the "tip radius forming line" around the tip central axis (Fig. 1b).

Both FE models were created using the static structural method with large displacements and the same material models: purely elastic material model for the indenter tip defined by its Young's modulus of 1140 GPa and Poisson's ratio of 0.07 (diamond), as used by Oliver and Pharr,[1] and the elastic–plastic material model with use of the power law approach, which is commonly used[2,10] for indented material. While the indenter tip is often assumed in the literature[2] to be completely rigid, the use of a deformable, purely elastic material model is justified by introducing a more realistic tip deformation (having impact on the tip contact area); the increase in model complexity is negligible. It was even shown by Mahdavi et al.[32] that the deviations introduced by the usage of an "equivalent rigid indenter approximation are by no means negligible for practical reason".[32] The power law material model of the indented body is described by the stress–strain relationships:

$$\sigma = E\varepsilon \text{ for } \sigma \leq \sigma_y, \quad (1)$$

$$\sigma = \sigma_y\left(1 + \frac{E}{\sigma_y}\varepsilon_p\right)^n \text{ for } \sigma \geq \sigma_y, \quad (2)$$

where $E$ is the elastic modulus, $\sigma_y$ the yield stress, $\varepsilon$ the total effective strain, $\varepsilon_p$ the plastic strain, and $n$ the strain-hardening exponent. The material model was implemented into the FE simulation using a Fortran-based material model subroutine (UMAT) by Martínez-Pañeda et al.[33]

The input parameters (material parameters defining their elastic and plastic behavior) of the simulations were inspired by Ref. 10 containing simulations from Ref. 2. The parameters were randomized to a certain amount to improve the generalization while still maintaining the comparability and to preserve the overall relationships of the parameters ($\frac{E}{\sigma_y}$). Tip radii were uniformly chosen between 50 nm and 1000 nm. The hardening parameters for the indented materials were between 0 and 0.5. This led to a rich dataset of inputs for the simulations, modeling numerous cases from both geometrical and material points of view. The data were analyzed to fulfil the same relationships as in Ref. 2 to avoid non-unique indentation curves, called mystical materials. These mystical materials have different materials parameters, but lead to practically non-distinguishable $P$-$h$ curves, as described by Chen et al.[34] Also, the "gold mine" of mystical materials was avoided ($\frac{E}{\sigma_y} \sim 100$ and small $n$).[34]

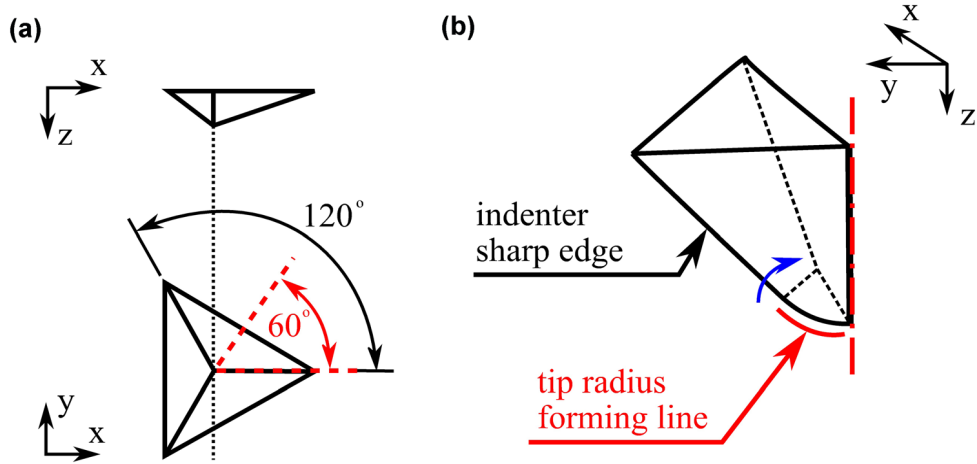

Fig. 1. (a) Schematic of the Berkovich tip and the symmetry planes (red dashed lines) used in the FE model. (b) Schematic of the introduction of the tip radius into the geometrical model, only a slice of the indenter tip is shown, whereas the red dash-dot line represents the central tip axis in the direction of indentation (Color figure online).

The indenter–sample contact was modeled as frictionless, which is commonly used in FE indentation studies.[2,35–38] The use of non-zero friction coefficients in the simulations of indentations is still under debate, partially because the determination of the friction coefficient in an actual experiment would be a non-trivial task. However, the choice of the coefficient of friction does not influence the obtained indentation curve; mainly, it is the pile-up or sink-in that are influenced.[39] Since the focus of this work is on the $P$-$h$ curves, a frictionless simulation is considered reasonable.

The maximum indentation depth was kept constant at 230 nm for both the 2D and 3D cases, and it was implemented in the FE model as a boundary condition prescribing the displacement in the direction perpendicular to the sample surface on the top of the indenter body. The whole simulation time span was divided into two sub-steps: the loading and unloading parts. In the first sub-step, the displacement of the indenter tip gradually increases from 0 nm to the maximum indentation depth. In the second sub-step, the indenter tip is moved from the maximum indentation depth back to its starting position. This way a quasi-static indent with both loading and unloading parts was modeled.

To obtain results with sufficient precision, a considerably fine FE mesh had to be used. However, the combination of the fine FE mesh and the relatively large deformations of the FE elements under the indenter tip can lead to numerical instability of the model (e.g., by excessive element distortions). To avoid such problems, an arbitrary Lagrangian–Eulerian adaptive mesh smoothing procedure was implemented in the presented models, leading to the reshaping of the elements in order to avoid mesh distortions and to increase the modeling efficiency with no impact on the numerical results.[40–43]

**Nanoindentation Experiments**

The nanoindentation experiments were performed using a TS77 Select Nanoindenter (Bruker/Hysitron) and a Berkovich diamond tip. The tip radius was characterized by confocal laser scanning microscopy (CLSM) before any indents were performed, as well as intermittently over the lifetime of the tip (when it was removed from the transducer). Additionally, the tip was periodically scanned over silicon spikes (TGT1 grid from NT-MDT) to create self-images of the tip, all of which were made using a 1-Hz scan rate, a load of 2–3 $\mu$N, and scan sizes of 5 $\mu$m and 2 $\mu$m. Self-images of the tip were made before and after every tip area function and frame compliance calibration, as well as after finishing a set of indents on various samples. Indents were made into materials including: fused silica, silicon, refractory metals and alloys (films and bulk), aluminum (film and bulk), copper (film and bulk), sapphire, WC-based materials (film and bulk), shale (geologic mineral), metallic glasses (film and bulk), polymeric materials, Mo-based oxides and nitrides (films), and biological materials. Open loop, displacement control, and mapping to various displacements were performed with the tip, as well as scanning for imaging. Tip wear is presented over nearly a year, and 29,694 indents were made with the same tip.

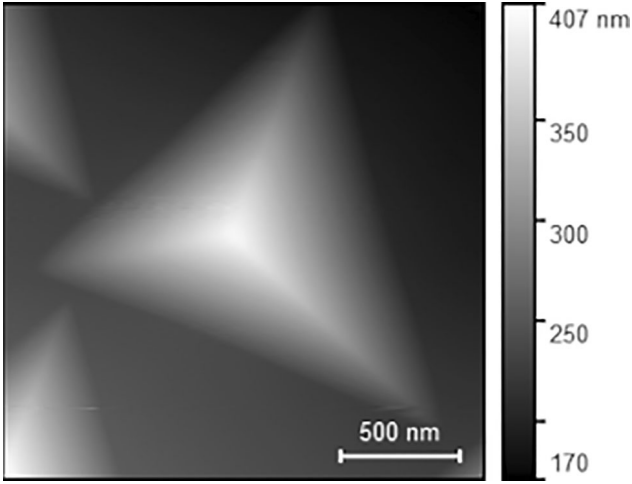

Fig. 2. Self-image of the Berkovich tip used with an initial measured radius of 420 nm.

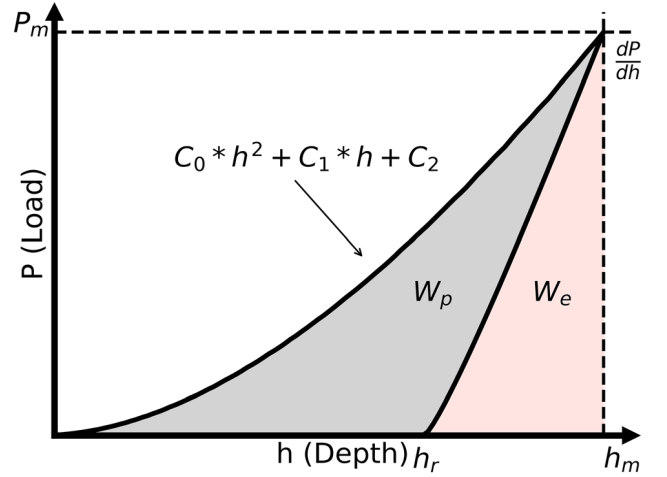

Fig. 3. Typical indentation curve obtained from simulation with elastic–plastic loading and elastic unloading.

The complexity of a direct measurement of a nanoindentation tip radius is still a task that should not be underestimated. Saringer et al.[19] showed that self-imaging over spikes is a reliable method for measuring the radius of an indentation tip, and that the method can also be used on CLSM images. This approach has been applied here using the 2-$\mu$m spike images to create an approximation of the tip radius of the indentation experiment, as seen in Fig. 2. The initial tip radius was measured as 420 nm. The tip radius was determined using the area function method as described in Ref. [19]. To ensure precise training of the surrogate model, the first five or the last five indents before or after a tip image were assumed to have the same tip radius as the image. This led to an experimental dataset consisting of 115 experimental data points for training and validation in different materials, and to 15 experimental data points for the testing set.

### Analyzing the Data

When analyzing the obtained $P$-$h$ curves, the curve has been divided into the elastic–plastic loading curve and the elastic unloading curve (Fig. 3). For sharp tips, the loading curve can classically be interpreted using Kick's law (Eq. 3); however, it was found not to precisely fit the real curve. Therefore, as proposed by Cheng and Cheng,[4] and based on the correction of Kick by Bernhartd,[44] the loading slope was instead approximated by a second-order polynomial (Eq. 4). Thus, a linear term ($C_1$) and a constant term ($C_2$) are included:

$$P = C * h^2, \qquad (3)$$

$$P = C_0 * h^2 + C_1 * h + C_2, \qquad (4)$$

where $P$ is the load, $h$ is the indentation depth, and $C, C_0, C_1,$ and $C_2$ are the fitted values. The initial slope of the elastic unloading curve can be fitted using the approach of Oliver and Pharr[1] to deliver the reduced elastic modulus after the tip's area function is calibrated. The total work, $W$, made of the elastic and plastic components, $W_e$ and $W_p$, respectively, can also be calculated using the respective areas under the $P$-$h$ curves. Li et al.[8] showed that the normalized curves $\left(\frac{P}{P_m}, \frac{h}{h_m}\right)$ are correlated to the tip rounding, and therefore the normalized features were also extracted (Table I). By analyzing the loading and unloading curves as described, the input values, called features, for the machine learning model are created.

### Machine Learning Methods

To accomplish data fusion, a residual-based multi-fidelity neural network (RMFNN) approach, as described by Lu et al.,[10] was used to bridge the fidelities from the 2D and 3D models to the actual experimental indentation data, with the aim of predicting the experimental tip radii. The setup consisted of two connected neural networks (NNs) that were trained with two datasets with different fidelities, as initially introduced by Meng et al.[17] The complex architecture was implemented in the Tensor Flow[45]-based DeepXDE library.[46] The architecture was trained in two steps:

- In the first step, the architecture was trained with the 2D simulation results as low-fidelity input data. The experimental data were split into a training and a validation set. The 3D simulation results combined with the experimental training data were used as the high-

Bridging Fidelities to Predict Nanoindentation Tip Radii Using Interpretable Deep Learning Models

**Table I. Features and calculation methods**

| Feature name | Calculation method |
| --- | --- |
| $C_0$ | Equation 4 |
| $C_1$ | Equation 4 |
| $C_2$ | Equation 4 |
| $C_{0,norm}$ | Equation 4 on normalized curve |
| $C_{1,norm}$ | Equation 4 on normalized curve |
| $C_{2,norm}$ | Equation 4 on normalized curve |
| $W$ | Area under loading curve (work) |
| $W_e$ | Area under elastic curve (elastic work) |
| $W_p$ | $W - W_e$ (Eq. 5) (plastic work) |
| work ratio | $\frac{W_p}{W}$ (Eq. 6) |
| $\frac{dP}{dH}$ | Fit of the initial unloading curve |
| $\frac{dP}{dH}$ norm | Fit of the initial normalized unloading curve |
| h-ratio | $\frac{h_m - h_r}{h_m}$ (Eq. 7) |

fidelity input. The experimental validation set was used for validation.

- In the second step, the model obtained from the first step was retrained using transfer learning, with the 3D simulation results as low-fidelity data and the training experiments as high-fidelity data. The same experimental validation set as in the first step was used for validation in the second step.

The approach utilizes the data and data combinations in an efficient way. Transfer learning is a state-of-the-art method to extend the predictive power of a model without having to start with a new, randomly initialized model. It is often used in modern deep learning libraries, mainly on vision and NLP tasks, such as in the fastai library,[47] to reduce the amount of data needed and to shorten computation times.[47] It utilizes pre-trained model weights as a base so that the learning process starts with information about the previous data (or even universal features), and is then adapted to the new data.[48] The approach is also fruitful for tabular data, as shown by Lu et al.[10] For tabular data, it is crucial that the features are fed in the same manner as before, otherwise there would be no benefit.[48]

The used RMFNN consists of two NNs, the low-fidelity NN having four layers and the high-fidelity having five layers. Both consist of 16 neurons per layer. They were trained for two steps with a learning rate of 0.01 for 5000 epochs using the Adam optimizer.[49] To avoid overfitting $L_2$ regularization, a size of 0.0005 was used.[50] The mean average percentage error (MAPE) was used to characterize the results,[51] defined as:

$$MAPE = \frac{1}{m} \sum_{i=1}^{m} \left| \frac{y_i - \widehat{y_i}}{y_i} \right|, \qquad (8)$$

where $y_i$ is the $i^{th}$ predicted value, $\widehat{y_i}$ is the ground truth, and $m$ is the total number of samples.

Four datasets were ultimately produced:

- 2D FE simulation dataset (401 data points)
  o Baseline training of the low-fidelity NN

- 3D FE simulation dataset (76 data points)
  o Baseline training of the high-fidelity NN
  o Transfer learning of low-fidelity NN in second step

- Experimental trainings/validation set (115 data points: 92 training 23 validation)
  o Baseline training of the high-fidelity NN
  o Validation of first model
  o Transfer learning of high-fidelity NN in second step
  o Validation of second model

- Experimental test set (15 data points)
  o Containing selected data completely hidden to the RMFNN during training to test the model after the second step

## RESULTS AND DISCUSSION

The training of the model was performed using transfer learning. To illustrate the differences between a model using transfer learning and a randomly initialized RMFNN, five-fold cross-validation was performed. In the first step, the model accomplished 15.1 ± 4.6 MAPE, and in the second (transfer learning) step 12.4 ± 3.8 MAPE (five-fold cross-validation). The randomly initialized RMFNN resulted in a MAPE of 13.17 ± 3.1, showing only minor differences from the pre-trained models. This means that the model could potentially also be trained only with the 3D and experimental datasets, discarding the 2D set completely. For further investigations, the transfer learned model was used. The test set was ultimately predicted with 16.1 MAPE, which can be seen as an improvement to the state-of-the-art predictions in which a calibration procedure always has to take place or the tip has to be scanned, as described by Saringer et al.[19] This is expected to enable future in situ predictions of the tip wear and its dependency on different materials, scanning protocols, or indentation depths/loads. The low error rate (i.e., MAPE value) for the test set and the five-fold cross-validation show that the model was able to predict the tip radius from the obtained $P$-$h$ curves.

The experimentally determined evolution of the tip radius is shown in Fig. 4. During the investigation time, the tip radius increases (blunts) and decreases (re-sharpens) while indenting several different materials using different indentation procedures and scanning. The high standard deviation for the first test set prediction (red triangle markers at indent 0) shows that the tip might be affected by every indent. The effect of a single indent appears to decrease with the increasing number of indents, as the standard deviation for the mean prediction also decreases with the increasing number of indents. Note that sharp indenters blunt faster compared to already blunt indenters, due to the higher pressures associated with sharp tips, as described by Nohava et al.[20] This could be part of the explanation for both the decrease in the standard deviation over time and the fast blunting of the tip during the first 5000 indents.

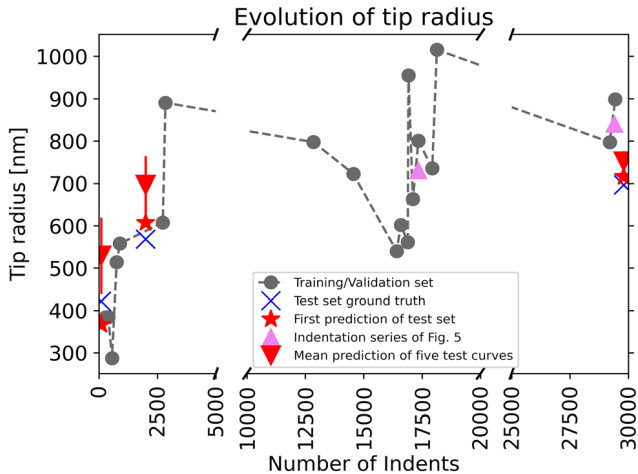

Fig. 4. Evolution of the tip radius over the accumulated number of indents, cut for long indentation periods (large maps) while still providing the same scale for all parts of the plot. The gray points (training/validation dataset) and blue Xs (test dataset) correspond to the tip radius measurements using the self-images. The red stars are the predictions of the very first curve after a tip radius measurement, while the red triangles are the mean predictions of the first five curves after the tip radius measurement and the respective standard deviation (Color figure online).

Predictions for tip wear of two different materials known to have isotropic behavior and similar elastic properties but different plastic behavior and hardness, are shown in Fig. 5. It depicts that the tip wear predictions differ depending on the materials and indentation depths. To be consistent with the training assumptions and mean predictions in Fig. 4, the running average over 5 experiments including standard deviation has been calculated and plotted. These two indentation series are used because tip radius measurements before and after indentation series were performed. Tip blunting occurs for both series according to the experimental results, marked by the upward-pointing triangles in Figs. 4 and 5. The first and last 5 indents for both series were used in the training process. This means that 10 indents are known to the model, with the evolution of the remaining 40 experiments predicted and presented in Fig. 5(a and b). What is remarkable is that the overall scatter of the predictions is low and that both plots follow clear trends with respect to the running average. This is further emphasized by the fact that the model has no information about the sequence itself, as neither the timestamp nor a cumulative number of indents is part of the training procedure. The fused quartz in Fig. 5a appears to blunt after an initial sharpening period, while the nanocrystalline aluminum leads to an overall gradual blunting of the tip (Fig. 5b). As a consequence of the evolution presented in Fig. 4, and the predictions in Fig. 5, more detailed experiments are planned to further prove the ability of the tip to sharpen.

As previously discussed, machine learning models can be benchmarked by the size of their prediction error (MAPE). To further analyze the seemingly black box model, SHAP[25] was used. Recall that SHAP allows for individual predictions based on cooperative game theory by approximating the Shapley values using local surrogate models.[52] The local explanations are consistent with the global interpretation.[52] The SHAP model agnostic interpretation, based solely on the inputs (features) and outputs (predictions) of the model, without the need for details about the complex model architecture, can be found. The data are presented in what is called a summary plot (Figs. 6 and 7), which combines feature importance and feature effects. Each point represents a single sample, while stacked or clustered points equal multiple data points with similar values. The order on the $y$-axis is determined by the importance of the feature, with the most important feature listed first. The colors correspond to the value of the feature itself. The values on the $x$-axis equal the SHAP values, where the zero line reflects the average model output. A positive SHAP value refers to an increase in the prediction value compared to the average model output and vice versa for negative SHAP values.

Figure 6 shows the described summary plot for the low-fidelity part of the RMFNN after transfer learning based on the validation set, and Fig. 7 shows the respective summary plot for the end prediction of the whole RMFNN. The differences in the order of the features shown in Figs. 6 and 7 can be attributed to differences between the simulated 3D data and the experimental data. These differences can be attributed to different levels of complexity of the data. All the features have a similar impact on both model predictions (a positive value triggers a positive reaction and vice versa), but different values are changing the model to a different degree. This can be seen in the color sequence for each individual feature, which remains the same in both diagrams (order-wise) despite the different impact on the model (width-wise). The similarities of the feature importance (order) and width indicates that the simulations show the same correlations as the experimental data.

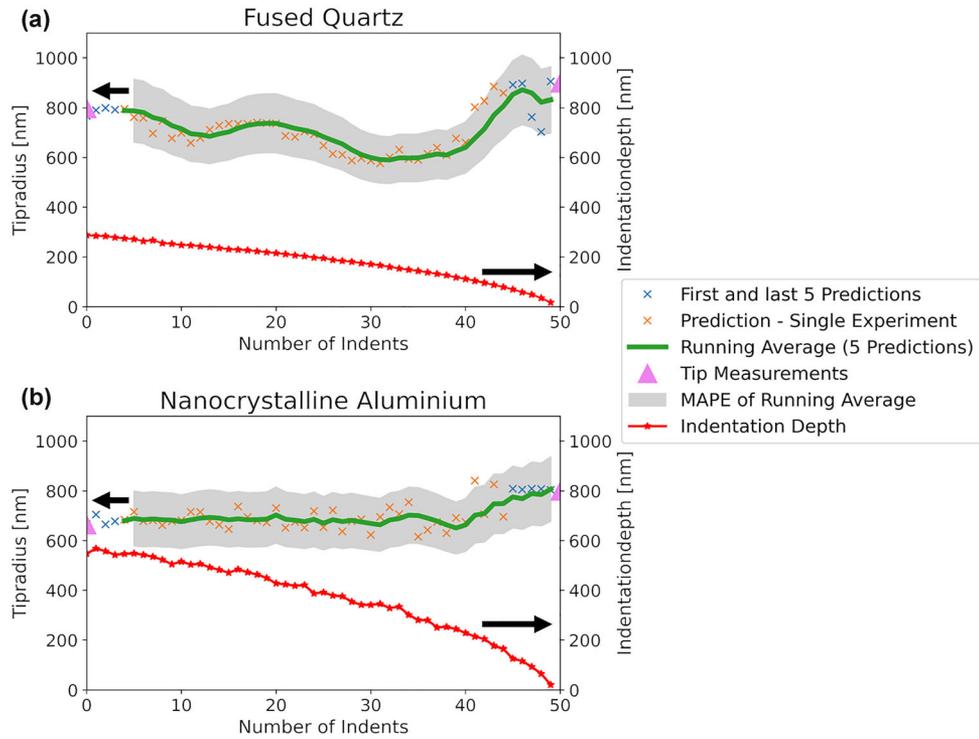

Fig. 5. Predictions of tip radius over 50 indents using different indentation depths including the running average over 5 predictions and its standard deviation for (a) fused quartz, corresponding to the first upward-pointing triangle in Fig. 4 (17,500 indents), and (b) nanocrystalline aluminum, corresponding to the second upward-pointing triangle in Fig. 4 (29,000 indents).

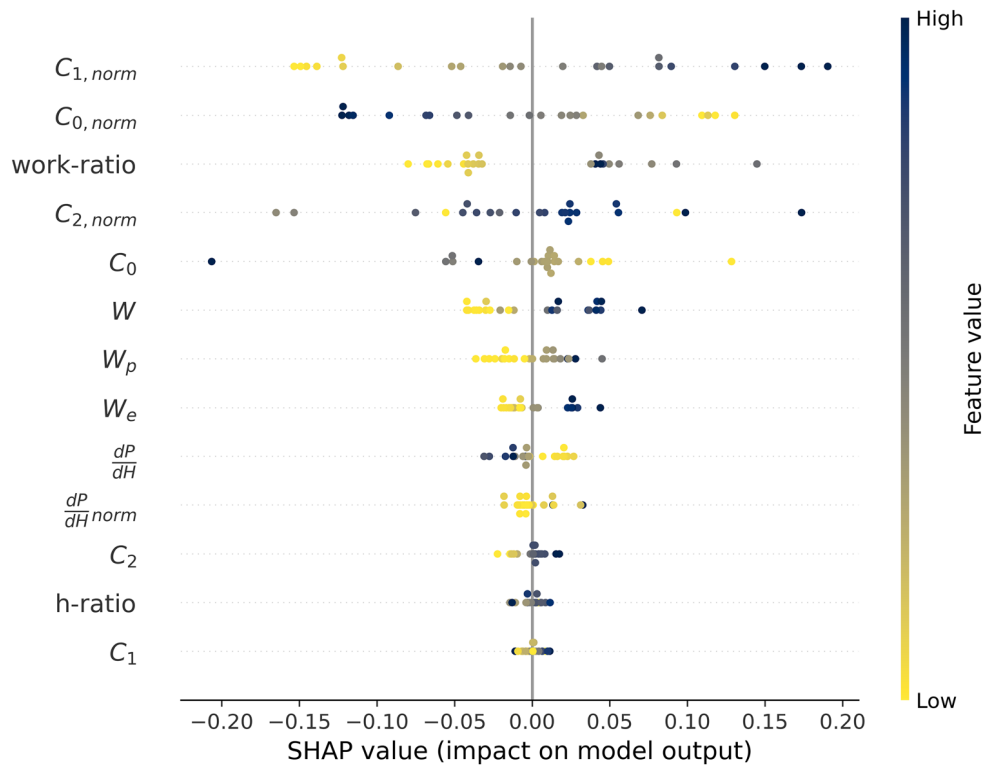

Fig. 6. SHAP summary plot of the low-fidelity output of the RMFNN evaluated on the experimental validation set.

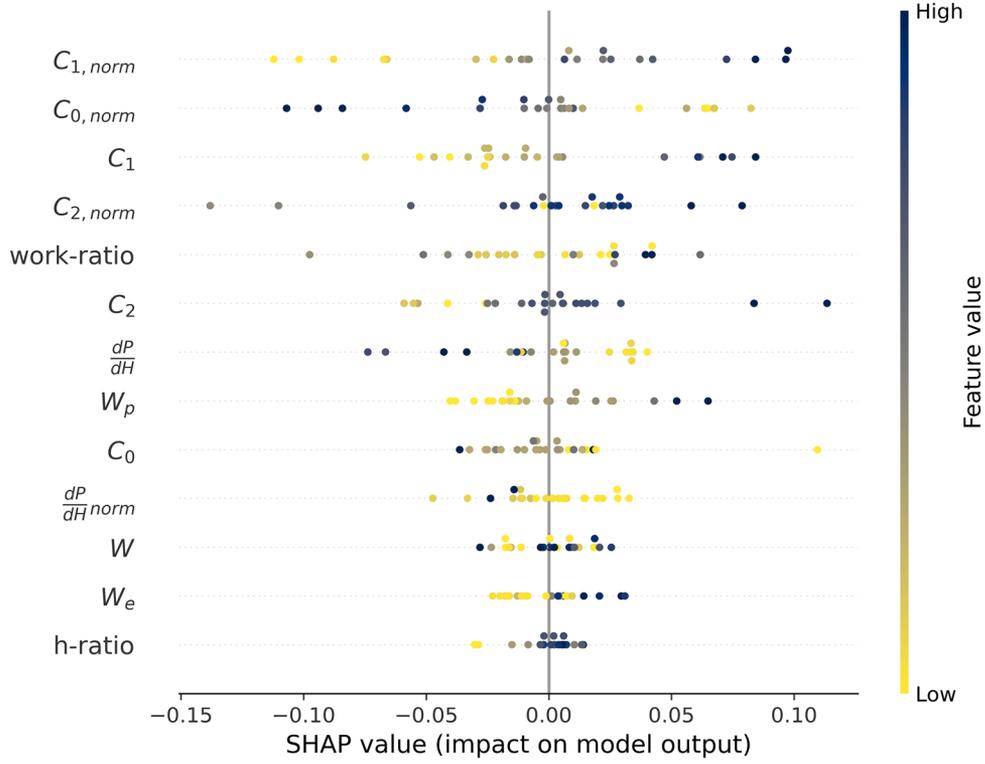

Fig. 7. SHAP summary plot of the end output of the RMFNN evaluated on the experimental validation set.

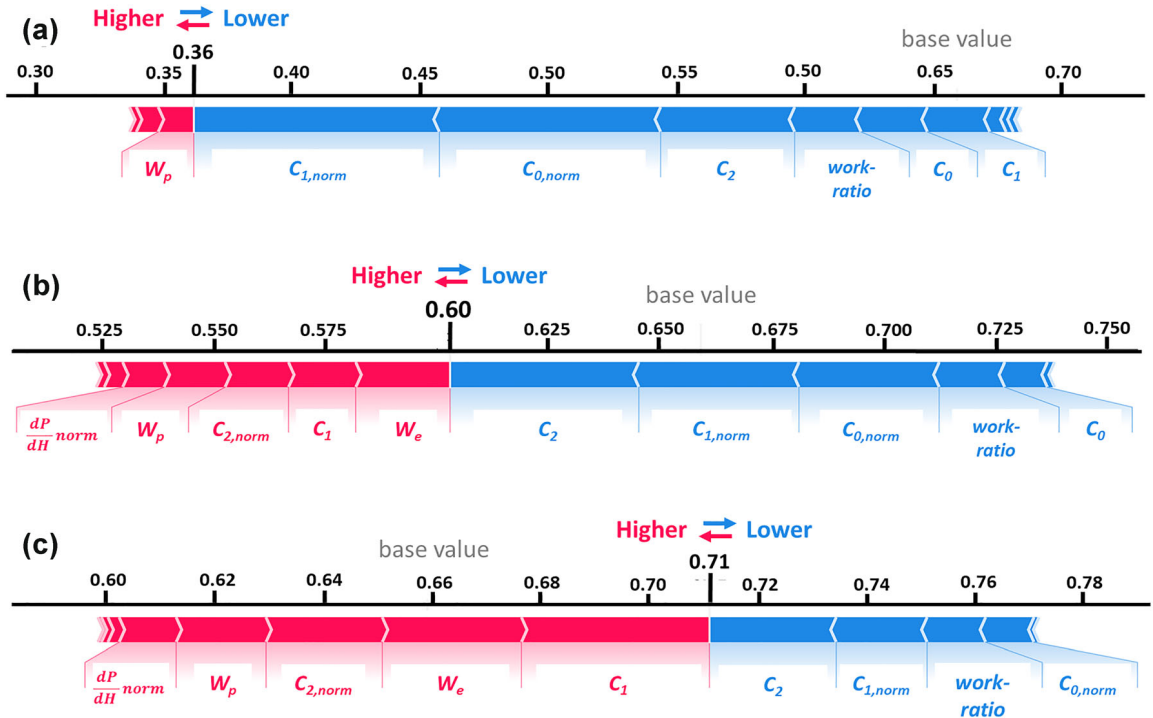

Fig. 8. (a–c) Local end prediction force plots on the test set data are based on the first of the five curves of the respective tip measurement (marked as red stars in Fig. 4).

When analyzing the feature importance (order of the features in Figs. 6 and 7), it can be seen that the features obtained from the normalized curvature are deemed the most important ($C_{1,\ norm}$ is listed first in both figures). This agrees well with the results of Li et al.,[8] who found that the curvature of

Bridging Fidelities to Predict Nanoindentation Tip Radii Using Interpretable Deep Learning Models

the normalized loading curve is related to the tip radius. The feature importance also indicates that fitting with Eq. 4, as proposed by Cheng et al.,[4] is a beneficial method to evaluate the tip radius using $P$-$h$ curves. The low-fidelity sub-model (Fig. 6) deems the loading part as the most important part of the model, as the top eight features are extracted from the loading curve. Also, for the end prediction (Fig. 7), the top six features are extracted from the loading curve. Additionally, $C_1$ and $C_2$ increase in importance from Figs. 6 and 7, which could be due to the fact that constant indentation depths are used in the simulations, whereas the experimental data has different indentation depths, thus influencing the loading curve depending on the actual depth. The analogies between the two figures show that the simulated and the experimental data evaluation are based on similar principles.

Figure 8 shows the detailed analysis of the predictions for the three first elements of the test set (red stars in Fig. 4). The features are represented as forces, pushing and pulling the prediction from the base value of the model. This can be created by taking every data point belonging to a single experiment (always one per feature), as seen in Figs. 6 and 7, and placing them into a force diagram. Figure 8 shows that a major part of the influences on the prediction originate from the loading segment of the indentation curve, namely features from the normalized curvature, curvature, $W$, $W_p$, and *work-ratio*. This plot can be used to monitor changes in prediction for single experiments by giving insights into the local prediction process.

## CONCLUSION

A nanoindentation tip was monitored over a period of nearly 30,000 indents, experiencing actual laboratory conditions using a vast amount of different materials and indentation procedures. During this time, the tip radius was regularly measured to quantify tip wear with a self-imaging method. It was shown that machine learning can be utilized to interpret this tip wear. Combining the knowledge of FE simulations, tip characterization, and experimental methods inside a surrogate model allowed the prediction and interpretation of the experimental tip wear, defined by changes in the tip radius measured by $P$-$h$ curves. Based on this research, future monitoring or calibration procedures for indentation experiments can be developed. The used approach allows not only characterization of the tip wear from indentation results with high precision but also enables further insight into the complex tasks inside a deep learning model. For the first time, complex multi-fidelity neural networks were analyzed and the decision-making process was examined from a game-theory perspective using SHAP. The study has also shown that interpretation of machine learning models can be a standard method in computational materials science in order to challenge the critique of non-interpretability of data-driven approaches. It will be crucial for scientists and engineers to continue to work as a control body, especially when it comes to safety-related applications. This approach is expected to allow the in situ study of tip wear, allowing the distinguishing between effects dependent on materials, scanning procedures, and indentation depths/loads, and will be shown in future work.

## ACKNOWLEDGEMENTS

The authors gratefully acknowledge the financial support under the scope of the COMET program within the K2 Center "Integrated Computational Material, Process and Product Engineering (IC-MPPE)" (Project No 859480). This program is supported by the Austrian Federal Ministries for Climate Action, Environment, Energy, Mobility, Innovation and Technology (BMK) and for Digital and Economic Affairs (BMDW), represented by the Austrian Research Funding Association (FFG), and the federal states of Styria, Upper Austria and Tyrol. Support by the Austrian Science Fund (FWF) under Grant No. P31140-N32 is acknowledged.

## FUNDING

Open access funding provided by Österreichische Akademie der Wissenschaften.

## DATA AVAILABILITY STATEMENT

All the used codes are uploaded to GitHub (https://github.com/materialsguy/Predict_Nanoindentation_Tip_Wear). All other experimental data are available upon reasonable request.

## CONFLICT OF INTEREST

The authors declare that they have no conflict of interest.

Bridging Fidelities to Predict Nanoindentation Tip Radii Using Interpretable Deep Learning Models